# A Framework for Android Based Shopping Mall Applications

Sajid Khan[1], Md Al Shayokh[2], Mahdi H. Miraz[3,4] and Moniruzzaman Bhuiyan[5]

[1]Samsung Research & Development Institute, Dhaka, Bangladesh
sajid.khan@samsung.com

[2]Department of ECE, Yildiz Technical University, Istanbul, Turkey
shayokh_2055r@yahoo.com

[3]Department of Computing, Glyndŵr University, Wrexham, UK
m.miraz@glyndwr.ac.uk

[4]Department of Computer Science & Software Engineering, University of Hail, KSA
m.miraz@uoh.edu.sa

[5]Institute of Information Technology, University of Dhaka, Bangladesh
mb@du.ac.bd

*Abstract-*Android is Google's latest open source software platform for mobile devices which has already attained enormous popularity. The purpose of this paper is to describe the development of mobile application for shopping mall using Android platform. A prototype was developed for the shoppers of Bashundhara Shopping Mall of Bangladesh. This prototype will serve as a framework for any such applications (apps). The paper presents a practical demonstration of how to integrate shops' information, such as names, categories, locations, descriptions, floor layout and so forth, with map module via an android application. A summary of survey results of the related literature and projects have also been included. Critical Evaluation of the prototype along with future research and development plan has been revealed. The paper will serve as a guideline for the researchers and developers to introduce and develop similar apps.

## I. INTRODUCTION

Mobile Phone users are no longer restricted to the basic predefined functionalities provided by the manufacturer and/or Mobile Operating Systems. Manufacturers now provide various development platforms and tools to support the third-party application developers. Not only that, some manufacturers even provide software markets such as Play Store by Google, where developers can sell and distribute their applications. Since the beginning of offering such opportunities, many different types of applications have been developed to cater the wide range of user demands. The concept of the Internet of Everything (IoE) has further geared up this phenomenon. Android has been playing a vital role in this regard. Our project is a part of such exertions of mobile application development using Android platform. We have developed the first iteration of the prototype to be used by shopping mall customers. Various Metadata of the participating shops of the Bashundhara Shopping Mall of Bangladesh were incorporated with an Indoor map of the Mall. This paper reports the finding of the initial evaluation of the prototype and will serve as a guideline for early developers and researchers interested into developing such apps.

## II. BACKGROUND AND METHODOLOGY

Mobile and cellular phone has gained huge popularity for its diverse use starting from making phone calls to sending SMS as well as act as a platform of running converged application and third party software. Due to such enormous diffusion of mobile devices and applications, mobile technologies have become a part of great and immense research. Android, introduced by Google, is a platform which is projected to be a complete software pack including operating system, middleware as well as core applications. In addition, it has an SDK (Software Development Kit) which provides the necessary tools for developing various applications in java platform [1,2].

Application markets such as Apple's App Store and Google's Play Store provide point and click access to hundreds of thousands of both paid and free applications. Such stores streamline software marketing, installation, and update therein creating low barriers to bring applications to market, and even lower barriers for users to obtain and use them. Design is very encouraging for developers and users of new applications as witnessed by the growing Android market. Android's application communication model further promotes the development of rich applications. Android developers can leverage existing data and services provided by other applications while still giving the impression of a single, seamless application [3]. For example, a restaurant review application can ask other applications to display the restaurant's website, provide a map with the restaurant's location and call the restaurant. This communication model reduces developers' burden and promotes functionality reuse. Android achieves this by dividing applications into components and providing a message passing system so that components can communicate within and across application boundaries [4]. Research has found that android apps are heavily being used to find out shopping mall locations, ongoing promotions and offers at different stores, map module etc. In developed country 44% [5] android users use shopping mall apps for their comfort. Developing countries are far legged behind in this regards but they have already started the journey to join the others.

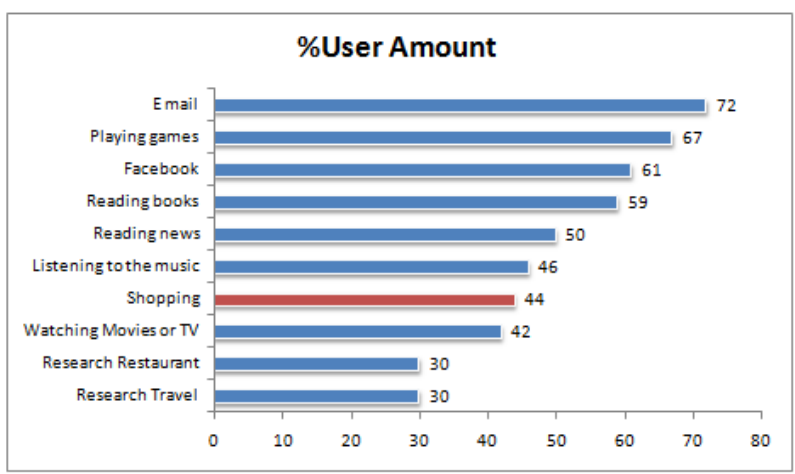

Figure 1. Percentage of global Android users for shopping purposes in 2013 [5]

For our project, we have opted for Android platform to develop the targeted app for shopping mall. Firstly, we conducted an extensive literature and project search to find out details of related works already done by other researchers and developers. We adopted the User Center Design (UCD) method to design and develop our first prototype. Participatory Design method, where representative(s) from the user groups take(s) part in the design process, was also used. The evaluation of the initial prototype has been presented here.

II. RELATED WORKS

The "Park Tower Shopping Mall" [6] App is one of the pioneer projects of this field. This was developed to be used by the customers of The Park Tower Shopping Mall which is located on main Shahrah-e-Ferdousi, Clifton in Karachi, Pakistan. This mobile application has the following functions:

1. The directory link function is to locate the address or the location of the shops.
2. Limited search functionality

3. The pop-up floor plans

The ShopZY [7] shopping Mall Mobile Application is developed for the shopping malls located in Bangalore and Ahmedabad. The Directory link function is to locate the address and location for the shops. The shops are divided into categories automatically. Consumers can use the search function on top of the screen to fasten the search process. The Layout link function is use to show the entire floor plans of the shopping mall. Other than these two main functions, there are other interactive links such as Favorites, Events, Promotions, Entertainment etc.

The official Simon Malls [8] App is developed for the Simon Property Group in USA. This is a must have shopping mall app for every savvy shoppers so that when they visit one of the Simon shopping malls out of 300+ malls nationwide, they can maximize their shopping experience. This app includes the visualization of the shopping mall map, checking out the latest promo offers as well as finding the desired shops etc.

The Dubai Shopping Mall [9] app is developed by Emaar Technologies. This is an app for the customers of Dubai Shopping mall to maximize their shopping experience. This app includes various features such as use interactive maps to help user pin point the location, showing detailed direction of desired place, real time update, latest events and promo etc.

The UAE Malls Combined [10] is an app developed my Mahmoud Khalil for UAE shopping malls. This app is an easy way to search and review of everything what an user/customer needs to know about his/her favorite shopping mall. This app has some unique features like completely designed smart search, navigating systems through the interactive maps, filter mall controller etc.

### III. PROPOSED MODEL AND ALGORITHM DESIGN

In this section, the technical model and developments algorithms have been discussed. A brief overview of the app, both technical and theoretical has also been presented here. For clarity of understanding, the section has been divided into the following sub-sections:

*A. Design Criteria*

This section reports the analysis of the system requirements for the app based on the user need. This has further helped us to identify the steps and procedures needed for the algorithm. Providing the floor layout has been identified as the primary requirement. In addition to that, displaying the layout of the selected floor has been pinpointed as to be a vital procedure while developing the algorithm for the app. Not only that, providing users with all the stores' information in the building comes as another prime fact due to its importance to the users. Besides, periodically data update service will be available to make user connected with the recent activity of the shopping mall. Moreover, category based search system will play a vital role for the popularity of this app. To add a new dimension to usability of the app, selected stores' interface could be added. Adding and letting the stores to update the promotion and offers will serve as a motivating factor for using the app as this will help: 1) the users to decide which shop to visit and 2) the stores to attract more customers. A use case diagram, as shown in fig. 2 has been presented to better the overall designing criteria.

Proc. of the Int. Conf. on eBusiness, eCommerce, eManagement, eLearning and eGovernance (IC5E 2014)

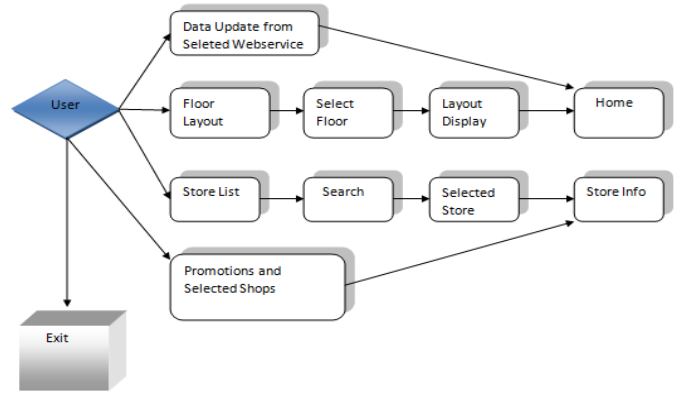

Figure 2. Use case diagram of the proposed App

Table-1 lists the findings of the requirement analysis and associates them with appropriate events to be triggers.

TABLE 1
DESIGN CRITERIA TABLE

| No | Requisite | Apply event |
|---|---|---|
| 1. | Providing Floor plans of the building | Layout of the Floor/Floor Layout |
| 2. | Providing selected floor layout | Select floor |
| 3. | Providing all the information about stores | Store list |
| 4. | Providing list of stores based on category | Category based search/search |
| 5. | Providing promotion offers and selected stores update | Promotion |
| 6. | Displaying the primary interface | Home |
| 7. | Terminating interface | Exit |

**B.** *Algorithm and Development of the app*

Considering those designing criteria, an efficient algorithm, as shown in fig. 3, has been developed for the system. At the beginning stage, data collection should be done and a total view of the shopping mall (Map view) should be generated. Next, all data will be categorized according to the floor plan and layout of the desired stores will be depicted. After that, category based search system should be developed and promotion features will be created.

Proc. of the Int. Conf. on eBusiness, eCommerce, eManagement, eLearning and eGovernance (IC5E 2014)

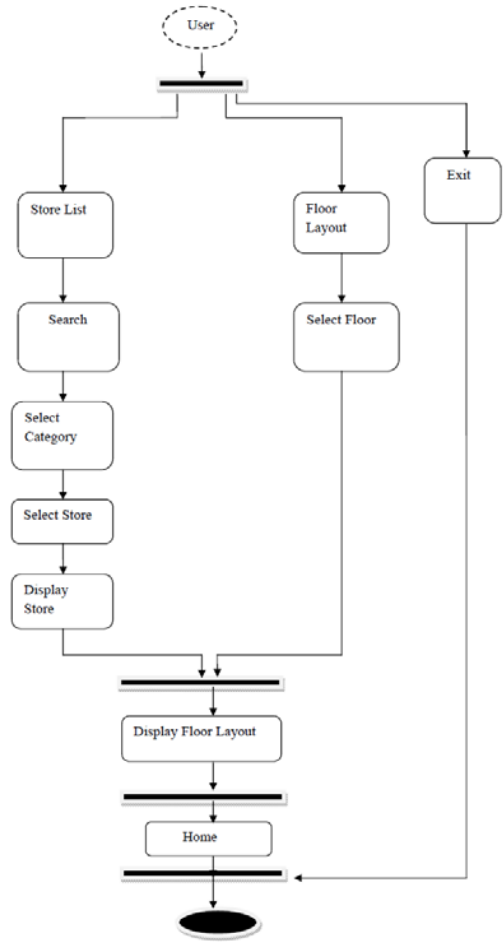

Figure 3. Algorithm development phase of proposed app

Fig. 4 presents some selected screen shots from the first iteration of the app. The initial user evaluation of the app provides positive feedback in terms of usability and usefulness. The results indicate that the app will be adopted and used by most of the customers having a smart phone or other compatible handheld devices. Once the final iteration is reached based on further detailed user evaluation, the app will be marketed for free for the users. However, for the shops to participate has to pay a monthly subscription to contribute to the development and maintenance process.

*Testing and Evaluation:*

The app was informally used, by a small number of users, for testing purposes only. The participants were mainly the shop owners and employees of the shopping mall. They are having previous experience of using smartphones. As they did not come across any such app before, they could not compare it with other similar apps as described in the Literature Review section of this paper. However, most of the participants were satisfied with the functionality of the app. The ability to navigate through the interactive maps was found to be the most prevalent feature. The participants were also pleased with other features as such search, filter, promotion and so forth. The app will be further enhanced based on the suggestions received and then a detailed usability test will be carried out. It will be subject to regular maintenance and updates.

Proc. of the Int. Conf. on eBusiness, eCommerce, eManagement, eLearning and eGovernance (IC5E 2014)

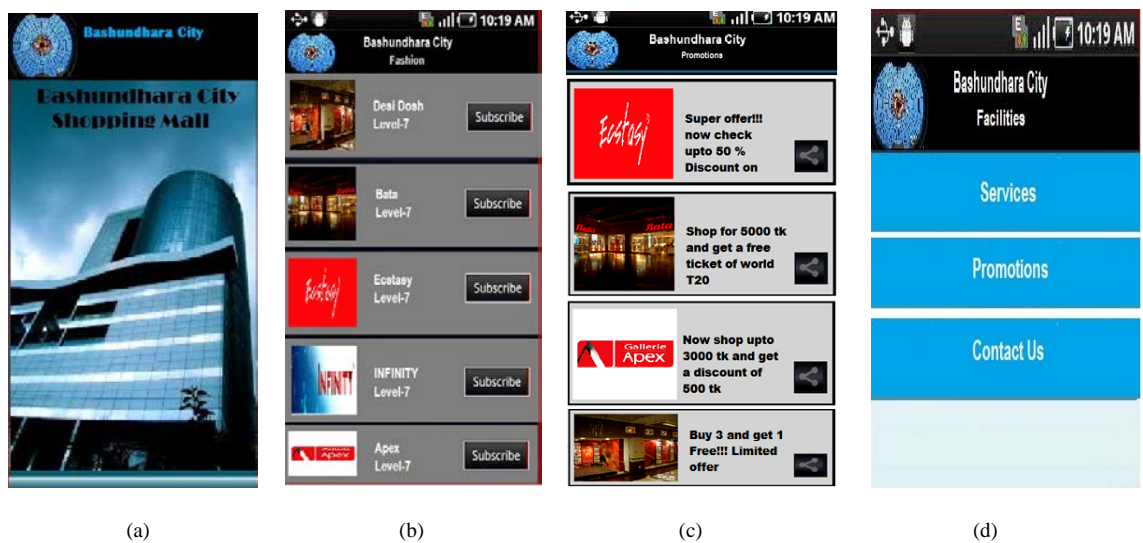

(a)　　　　　　　　(b)　　　　　　　　(c)　　　　　　　　(d)

Figure 4. An overview of the proposed Android App

## IV. CONCLUSION

This Paper presents an overview of how to develop an Android based shopping mall app from the scratch. Processes of doing requirement analysis using UCD and participatory design methods as well as generating development phase algorithm have been discussed. The initial prototype was developed and evaluated based on the requirement analysis. The initial evaluation of the prototype provided very encouraging results. However, to develop a marketable version of the app, the prototype has to undergo numerous further iterations until high user satisfaction is achieved. A detailed usability evaluation has to be carried out. The prototype developed is expected to serve as a framework for any such applications.